\documentclass[12pt,a4paper]{article}

\usepackage[latin1]{inputenc}

\usepackage{amsfonts}
\usepackage{amssymb}
\usepackage{amsthm}
\usepackage{latexsym}
\usepackage{fancyhdr}
\usepackage{psfrag}

\newtheorem{defin}{\textbf{Definition}}

\newcommand{\C}{\mathbb{C}}
\newcommand{\R}{\mathbb{R}}

\newcommand{\ot}{\theta}
\newcommand{\ke}[1]{|#1\rangle}

\newcommand{\bk}[2]{\langle #1|#2\rangle}
\newcommand{\vv}[1]{``#1''}
\newcommand{\sech}{\textrm{sech}}

\title{Robust Steering of n-level Quantum Systems}
\author{Francesco Ticozzi\thanks{Dipartimento di Ingegneria dell'Informazione,
Universit\`a di Padova, via Gradenigo 6/B, 35131 Padova, Italy
({\tt francescoticozzi@libero.it}).} \and Augusto
Ferrante\thanks{Dipartimento di Ingegneria dell'Informazione,
Universit\`a di Padova, via Gradenigo 6/B, 35131 Padova, Italy
({\tt augusto@dei.unipd.it}).}\and Michele
Pavon\thanks{Dipartimento di Matematica Pura e Applicata,
Universit\`a di  Padova and ISIB-CNR,
   via Belzoni 7,
35131 Padova, Italy ({\tt pavon@math.unipd.it}).
}  }
\date{\today}

\begin{document}

\maketitle

\abstract Robust open-loop steering of a finite-dimensional
quantum system is a central problem in a growing number of
applications of information engineering. In the present paper, we
reformulate the problem in the classical control-theoretic setting, and
provide a precise definition of {\em robustness} of the control
strategy. We then discuss and
compare some significant problems from NMR in the light of the given
definition. We obtain quantitative results that are consistent with the
qualitative ones available in the physics literature.

\section{Introduction}
We consider an isolated n-dimensional quantum system with time
evolution described by the following Schr\"{o}dinger equation:
\begin{eqnarray}\label{model}
i\hbar\ke{\dot{\psi}(t)}&=&H(t)\ke{\psi}.
\end{eqnarray}
Here $\ke{\psi(t)}$ is a vector of unit norm in $\mathbb{C}^n$
representing the state of the system at time $t$. The unitary time
evolution of the system is governed by the system Hamiltonian:
\begin{equation}\label{Htot}
H(t)=H_0+\sum_{j=1}^{m}H_j u_j(\ot,t).
\end{equation}
The \emph{internal Hamiltonian} $H_0\in\C^{n\times n}$ is an
Hermitian matrix describing the free evolution of the system. The
\emph{control Hamiltonian} $$H_c(t)=\sum_{j=1}^{m}H_j
u_j(\ot,t),$$ where $H_j\in\C^{n\times n}$ are also Hermitian
matrices, accounts for the effects of the control inputs
$u_1(\ot,t),...,u_m(\ot,t)$ on the dynamics of the system. We
assume that these control functions depend on a finite number of
real parameters $\ot=(\theta_1,...,\theta_p),\,
\theta_k\in\mathcal{T}$, with $\mathcal{T}$ being an open set in
$\R^p$. This kind of assumption is reasonable if we think of the
small set of parameters we can control in an experimental setting.

We consider the problem of steering the system from a given
initial state $\ke{\psi_0}=\ke{\psi(0)}$ to a final state
$\ke{\psi_1}$, where $\ke{\psi_0}$ and $\ke{\psi_1}$ are unit
vectors in $\C^n$. We assume that the transition occurs (at t=T)
when $\theta=\theta^*$ which we take as \vv{nominal} value of the
parameters. Clearly, if $\ot\neq\ot^*$, the transition will, in
general, not occur.

It is convenient to introduce the \emph{error probability} for
each control strategy. Consider the  normalized final state for
the time evolution, $\ke{{\psi}(T,\theta)}$. It can be written as
$\ke{{\psi}(T,\theta)}=\bk{\psi_1}{{\psi}(T,\theta)}\ke{\psi_1}+\ke{\psi^\bot(\ot,T)}$
with $\ke{\psi^\bot(\ot,T)}$ orthogonal to $\ke{\psi_1}$. If we
imagine to perform a discrete measure\footnote{Quantum measure
fundamental postulates can be founded in standard quantum
mechanics textbooks, see e.g. \cite{sakurai},\cite{messiah} or
\cite{cohen}.} on an observable that has $\ke{\psi_1}$ as
eigenstate,  the probability to obtain the eigenvalue associated
$\ke{\psi_1}$(that corresponds to the probability of finding the
system in $\ke{\psi_1}$ immediately after the measure) is:
$P_{\ke{\psi_1}}=|\bk{\psi_1}{{\psi}(T,\theta)}|^2.$ Then the
\emph{error probability} corresponding to the value $\ot$ is:
\begin{eqnarray}\label{Perr}
P_{err}(T,\ot)&=&1-|\bk{\psi_1}{{\psi}(T,\theta)}|^2
\nonumber\\
&=&\bk{\psi^\bot(\ot,T)}{\psi^\bot(\ot,T)},
\end{eqnarray}
thanks to the fact that $\ke{{\psi}(T,\theta)}$ is normalized. By
assumption, we have $P_{err}(T,\ot^*)=0$.

\section{Robustness of the control strategy}\label{robustness}
In the quantum control field, the expression \vv{robustness of the
control strategy} means that the control performance is
insensible  with respect to errors in the control implementation. In
\cite{optimalap}, a control strategy  is considered robust
\vv{\emph{if significant local changes in the amplitude and the
form of the pulse and of the chirp do not change significantly the
final transfer probability}.} The pulse and the chirp, in the
setting described in Section \ref{examples}, are the system inputs
parameters. A quantitative definition of robustness is, however,
missing.

The contribution of the present paper is to provide such a
quantitative definition reformulating the problem as a robust
control-theoretic problem, and then to analyze the robustness properties
of some significant strategies considered in the relevant
literature. In terms of our model, this concept of robustness can
be qualitatively formulated as follows: A  control strategy is
robust when, for values of the parameters $\ot$ different
from the nominal ones, the final state $\ke{{\psi}(T,\theta)}$ is close
to the desired one $\ke{\psi_1}$. This robustness request is
satisfied if  $P_{err}(T,\ot)$ is small in the parameter set
$\mathcal{T}$.

In classical control theory, plant uncertainty is described by a
set $\mathcal{P}$ of possible plants \cite{doyle-feedback}. This
uncertainty can be either structured (parametrized by a finite
number of scalar parameters or a discrete set of plants) or
unstructured (disc-like uncertainty). A controller is said to be
robust with respect to some property if this property holds for
every plant in $\mathcal{P}$.
It is quite simple to reformulate our problem as a particular case
of structured-like classical robustness problem. First, we
notice that our quantum \vv{plant} is determined by the matrices
$(H_0,H_1,...,H_m)$. These matrices determine the system
Hamiltonian (\ref{Htot}), given the control strategy.
Let $P_0=(H_0,H_1,...,H_m)$ be our nominal plant.

As in \cite{viola-robustness}, we can transfer the
uncertainty from the control parameters to the internal
Hamiltonian. In fact, by defining $\delta
u_i(\theta)=u_i(\theta)-u_i(\theta^*)$ we can write:
\begin{eqnarray}\label{Hu}
{H}(t)&=&H_0+\sum_{i=1}^m H_i \left(u_i(\ot^*)+\delta
u_i(\ot)\right)\nonumber\\
&=&H_0+\sum_{i=1}^m H_i \delta u_i(\ot)+\sum_{i=1}^m H_i
u_i(\ot^*)\nonumber\\
&=&\left(H_0+\Delta H_u(\ot)\right)+\sum_{i=1}^m H_i u_i(\ot^*).
\end{eqnarray}where $\Delta H_u(\ot)=\sum_{i=1}^N H_i \delta
u_i(\ot)$. Such a cosmetic transformation shows that our control
strategy uncertainty can be seen as a particular case of the plant
uncertainty (with control inputs $u_i(\ot^*)$). The plant set
$\mathcal{P}$ is here given by:
$$ \mathcal{P}=\{(H_0+\Delta
H_u(\ot),H_1,...,H_m)|\theta\in\mathcal{T}\}.$$
The property we are interested in, as e.g. in \cite{optimalap}, is
the error probability. We require this probability not to exceed a
fixed threshold $\epsilon\in[0,1)$ at a given $T$.
All the ingredients of a classical robustness problem have now been
specified. Introduce the \emph{$\epsilon$-robustness set}
$\mathcal{R}_\epsilon$ as
\begin{equation}\mathcal{R}_\epsilon=
\{\ot\in\mathcal{T}|P_{err}(\ot,T)\leq\epsilon\}.\end{equation}
We give the following definition.
\begin{defin} A control strategy
$\{u_1(\ot^*,t),...,u_m(\ot^*,t)\},\, t\in[0,T]$ is
\emph{$\epsilon$-robust} with respect to parameters uncertainty
if:
\begin{equation}\mathcal{R}_\epsilon=\mathcal{T}.\end{equation}
\end{defin}
\noindent Notice that only the 0-robustness case ensures us an
exact steering of the system state for all
$\ot\in\mathcal{T}$.

\section{Some applications}\label{examples}

In this section we analyze, in the light of the above definition,
the robustness of some prototype examples. We will compare our
results with qualitative observations and robustness claims in the
relevant literature. To do so, we introduce a particular form of
(\ref{model}) frequently used in NMR (Nuclear Magnetic Resonance)
quantum control problems. We consider a two level quantum system,
and the associated a bi-dimensional Hilbert space. The time
evolution is described by a scaled time Schr\"{o}dinger equation in
the form:

\begin{equation}\label{smodel}
i\hbar\frac{\partial}{\partial s}\ke{{\psi}(s)}=TH(s)\ke{\psi(s)},
\end{equation}
where $s=t/T$ and
$$H(s)=\left(%
\begin{array}{cc}
    -\Delta(s) & \Omega(s) \\
    \Omega(s) & \Delta(s) \\
\end{array}%
\right),$$
is represented in the canonical (\emph{diabatic}) base.
The control functions:
$$\left\{\begin{array}{l}
   \Delta(s)=\Delta_0\Phi(s)\\
   \Omega(s)=\Omega_0\Lambda(s) \\
\end{array}\right.,$$
are the inputs, with $\Phi(s)$,\,$\Lambda(s)$ fixed envelops and
$\Delta_0$, $\Omega_0\in\R^+$ amplitude parameters. In this
picture we have $\Delta(s)=u_1(s,\Delta_0)$ and
$\Omega(s)=u_2(s,\Omega_0)$. Thus
$\ot=(\theta_1,\theta_2)=(\Delta_0,\Omega_0)$ are the parameters
we are interested in. In the contest of particle-laser field
interaction and the RWA (Rotating Wave Approximation
\cite{alleneberly}), these functions depend on the chirp
(\emph{detuning})  and the amplitude (\emph{time-dependent Rabi
frequency}) of the active pulse\footnote{To find some detailed
information about the physical meaning of these parameters and
about the resonance phenomenon see
\cite{vitanov-review},\cite{NMR}.}. This model can be seen as a
particular case of model (\ref{model}), and it is suitable to
describe control techniques based both on magnetic resonance and
adiabatic passage. At each time the unitary transformation
\begin{equation}
U(s)=\left(%
\begin{array}{cc}
    \cos\theta(s) & -\sin\theta(s)\\
    \sin\theta(s)& \cos\theta(s) \\
\end{array}%
\right),\end{equation} with
$\theta(s)=\frac{1}{2}\arctan\left(\Omega(s)/\Delta(s)\right)$,
diagonalizes the Hamiltonian\begin{equation}
U^\dag(s)H(s)U(s)=\left(%
\begin{array}{cc}
    \varepsilon(s) & 0 \\
    0 & -\varepsilon(s) \\
\end{array}%
\right)=D(s).\end{equation}
Here
\begin{equation}\label{E}\varepsilon(s)=\sqrt{\Delta^2(s)+\Omega^2(s)}.
\end{equation}
is the energy eigenvalue.
Applying $U(s)$ as a time dependent change of basis and defining
$\phi$, we obtain:
\begin{eqnarray}
i\hbar \frac{\partial\ke{\phi(s)}}{\partial s}&=&
\left[TD(s)-iU^\dag(s)\frac{\partial}{\partial
s}U(s)\right]\ke{\phi(s)}\nonumber\\
&=&\left(%
\begin{array}{cc}
    T\varepsilon(s) & i\gamma(s) \\
    i\gamma(s) & -T\varepsilon(s) \\
\end{array}%
\right)\ke{\phi(s)}.\end{eqnarray} The new basis vectors are
called \emph{adiabatic} states. In the adiabatic limit,
$T\rightarrow\infty$, the $\gamma(s)$ terms can be neglected, as
shown by the adiabatic approximation theory\cite{messiah}.

In the following subsections, we will investigate the robustness properties of
different control strategies in a typical steering problem, the
\textit{state-flipping}: Transfer the system state from one basis
vector to the other. The standard resonance technique and two
adiabatic models will be discussed and compared.

\subsection{Magnetic Resonance}
A simple way to obtain such a transfer is by using the magnetic
resonance phenomena: Under properly tailored oscillating fields,
the state vectors rotate between the two basis states
\cite{vitanov-review},\cite{NMR}. This kind of effect can be
generated by the following fields-control functions:
\begin{equation}
\left\{ \begin{array}{l}
    \Delta(s)=0 \\
    \Omega(s)=\Omega_0 \Lambda(s), \\
\end{array}\right.\end{equation}
where $\Lambda(s)$ is the $\Omega$-pulse envelope. This
parametrization, and some easy calculations \cite{sakurai}, lead
to the following expression for the error probability:
\begin{equation}
P_{err}(T,\Omega_0,A_\Lambda)=\cos^2\left[T\Omega_0\int_{s_i}^{s_f}\Lambda(s)ds\right]=\cos^2\Omega_0TA_\Lambda,
\end{equation}
with $A_\Lambda$ the $\Omega$-pulse area.

This probability is equal to zero for:
$$\Omega_{0,k}^*=\frac{(k+\frac{1}{2})\pi}{TA_\Lambda},\,
k=0,1,2,... ,$$ or, equivalently, for:
$$A_\Lambda^*=\frac{(k+\frac{1}{2})\pi}{T\Omega_0},\,
k=0,1,2,... .$$ Thus, $P_{err}(T,\Omega_0,A_\Lambda)=0$ in a
family of hyperbolas parameterized in $k$ (zero-measure set in the
parameters space). Consider
$\mathcal{T}=[\Omega_0^*-\beta,\Omega_0^*+\beta]\times[A_\Lambda^*-\sigma,A_\Lambda^*+\sigma]$,
a common setting if we are working with nominal values and error
intervals, with $\beta,\sigma$ such that $\sin{2T\Omega_0
A_\Lambda}$ is monotone in every direction. Then, the maximum
absolute value for the error probability in $\mathcal{T}$ is:
\begin{equation}
P_{max}=\cos^2T\bar{\Omega}\bar{A}_\Lambda,
\end{equation}
where $\bar{A}_\Lambda=A_\Lambda^*+\sigma$ and
$\bar{\Omega}=\Omega_0^*+\beta$. Then $\mathcal{T}$ is
$\bar{\epsilon}$-robust, with $\bar{\epsilon}=P_{max}$.

\noindent According to qualitative evaluation found, the magnetic
resonance strategy doesn't seem to ensure enough insensibility
towards errors in control implementation and can be sensitively
improved by adiabatic passage techniques.

\subsection{Landau-Zener Model}
The Landau-Zener model is one of the simplest choice of control
function leading to an adiabatic transition. We will consider:
\begin{equation}
\left\{ \begin{array}{l}
    \Delta(s)=\frac{\Delta_0^2}{T} s \\
    \Omega(s)=\Omega_0  \\
\end{array}\right.,\end{equation}
the \emph{detuning} varies linearly with a zero crossing, while
the Rabi frequency is maintained constant. For $s=0$ we have a
minimum in the difference between energy levels that leads to a
state inversion if the evolution satisfies  the condition needed
for the adiabatic approximation. The error probability is
estimated with the Landau-Zener formula:
\begin{equation}\label{landauzenerformula}
P_{err}(T,\Omega_0,\Delta_0)\approx e^{-\pi
T\frac{\Omega_0^2}{\Delta_0^2}}.
\end{equation}
This probability goes to zero in the adiabatic limit
$T\rightarrow\infty$ for any choice of $\Omega\neq 0$ and . Thus,
the robustness set for this strategy is the whole open first
quadrant without its boundary ($\Omega_0=0,\Delta_0=0$).
The advantages given by this adiabatic technique are evident, as
long (\ref{landauzenerformula}) estimates correctly the error
probability. Even if the $T\rightarrow\infty$ condition is not
realizable, we can take a $T$ large enough to maintain $P_{err}$
arbitrary small for (almost-)every parametrization of the control
strategy. We will call this behavior \emph{intrinsically robust}.

\subsection{Allen-Eberly Model}
We now analyze an adiabatic control strategy more complex than the
previous one. The Allen-Eberly \cite{alleneberly} parametrization
allows to obtain an exact expression for the error probability
and, in the $\Omega_0=\Delta_0$ case, forces the state time
evolution along the energy \emph{level lines}, maintaining
$\varepsilon(\Omega(s),\Delta(s))=c $, $c$ constant for every $s$
\cite{optimalap}. This kind of choice leads to good results in
terms of error probability even quite far from the ideal
$T\rightarrow\infty$ condition, as we are going to show. In terms
of control functions, we consider:
\begin{equation}
\left\{ \begin{array}{l}
    \Delta(s)= \Delta_0\sqrt{1-\sech^2(s)}=\Delta_0\tanh(s)\\
    \Omega(s)=\Omega_0 \sech(s) .\\
\end{array}\right. \end{equation}
Then, the exact expression for the error probability is:
\begin{equation}
P_{err}(T,\Omega_0,\Delta_0)=\cosh^2\left(\pi
T\sqrt{\Delta_0^2-\Omega_0^2}\right)\sech^2\left(\pi\Delta_0
T\right),
\end{equation}
for every regime, adiabatic or not. We can notice that, for large
$T$ and for $\Delta_0\geq\Omega_0$, the error probability can be
bounded by:
\begin{equation}
P_{err}(T,\Omega_0,\Delta_0)\leq 4 e^{-2\pi
T(\Delta_0-\sqrt{\Delta_0^2-\Omega_0^2})}.
\end{equation}
Thus, for every $\Delta_0$ and $\Omega_0$, $\Delta_0\geq\Omega_0$,
the error probability decreases exponentially to zero in the
adiabatic limit. The best choice for the parameters values is to
take the largest $\Delta_0=\Omega_0$.
In the case $\Omega_0>\Delta_0$, the error probability becomes:
\begin{equation}
P_{err}(T,\Omega_0,\Delta_0)=\cos^2\left(\pi
T\sqrt{\Omega_0^2-\Delta_0^2}\right)\sech^2\left(\pi\Delta_0
T\right).
\end{equation}
This expression tends to zero with dumped oscillations, due to the term
$\cos^2\left(\pi T\sqrt{\Omega_0^2-\Delta_0^2}\right)$. Again,
larger $\Delta_0$ make $P_{err}$ converge faster.
Thus, for each fixed $\epsilon$, we can compute a $T_\epsilon$
such that the error probability
$P_{err}(T,\Omega_0,\Delta_0)<\epsilon$ for every $T>T_\epsilon$.
Indeed,  it is easy to see that
$$T_\epsilon
=\max\{-\frac{\ln{\frac{\epsilon}{4}}}{2\pi(\Delta_0-\sqrt{\Delta_0^2-\Omega_0^2})},
\ - \frac{\ln{\epsilon}}{2\pi \Delta_0}\}.$$ This control strategy
is therefore \emph{intrinsically robust} for $T$ sufficiently
large. According to the Landau-Zener case, every choice of
$\Omega_0\neq0$ and $\Delta_0\neq0$ drives the system to the
target state. The level line condition $(\Delta_0=\Omega_0)$ and
large $\Omega_0$ give faster convergence to the desired state.

\section{Discussion}
Comparing the results, the advantages given by the adiabatic
strategies are evident. They can be effectively used, however,
when the transfer time is not critical: Their intrinsic robustness
is exhibited only with a large time.

The examples analyzed are also treated in \cite{optimalap} to
illustrate that control strategies based on the level lines are
optimal for adiabatic population transfer (the level line
strategies minimize the error probability and corresponds to the
minimum pulse area). In \cite{optimalap}, robustness of the
control is also taken in account: Contour plots of error
probability with respect to parameters variations are obtained
thanks to numerical simulations for the system evolution.
Different strategies are qualitatively compared. It is shown that
the simple resonance case generates larger error probability than
the adiabatic optimal techniques, once a parameter variation is
fixed.

Here we have given a formal definition of the robustness property,
reformulating the problem in the control theoretical setting. We
have obtained quantitative results consistent to the qualitative
ones just described, and we have provided an analysis tool useful
to evaluate and compare robustness behavior of different
strategies. From a control theoretic viewpoint, we have analyzed
a specific robustness problem for open-loop control of a bilinear
system.

\bibliographystyle{plain}

\begin{thebibliography}{1}

\bibitem{alleneberly}
L.~Allen and J.H. 
Eberly.
\newblock {\em Optical Resonance and Two-Level 
Atoms}.
\newblock Dover, 1987.

\bibitem{cohen}
C.~Cohen-Tannoudji, 
B.~Diu, and F.~Laloe.
\newblock {\em Quantum Mechanics}.
\newblock 
Wiley \& Sons, 1977.

\bibitem{doyle-feedback}
J.C. Doyle, A.~B. 
Francis, and A.R. Tannenbaum.
\newblock {\em Feedback Control 
Theory}.
\newblock Macmillan Publishing Company, 
1992.

\bibitem{NMR}
M.~Mehring and V.~A. Weberrun.
\newblock {\em 
Object-oriented magnetic resonance}.
\newblock Academic press, 
2001.

\bibitem{messiah}
A.~Messiah.
\newblock {\em Quantum 
Mechanics}.
\newblock Dover, 1999.

\bibitem{sakurai}
J.J. 
Sakurai.
\newblock {\em Modern Quantum Mechanics}.
\newblock 
Addison-Wesley, revised edition, 
1994.

\bibitem{optimalap}
S.~Thomas, S.~Guerin, and H.R. 
Jauslin.
\newblock Optimization of population transfer by adiabatic 
passage.
\newblock {\em Phisical Review A}, 65:023409--1,5, 
2002.

\bibitem{viola-robustness}
L.~Viola and E.~Knill.
\newblock 
Robust dynamical decoupling with bounded controls.
\newblock {\em 
preprint available at Los Alamos on-line archive}, 
2002.

\bibitem{vitanov-review}
N.V. Vitanov, T.~Halfmann, B.W. 
Shore, and K.~Bergmann.
\newblock Laser-induced population transfer 
by adiabatic passage techniques.
\newblock {\em Annu.Rev.Phys.Chem.}, 
52:763--809, 2001.

\end{thebibliography}

\end{document}